\documentclass[twocolumn,prd,floatfix,preprintnumbers,a4paper,nofootinbib,superscriptaddress,groupedaddress]{revtex4-1}
\usepackage[utf8]{inputenc}
\usepackage{amsmath,amssymb}
\usepackage{amsfonts}
\usepackage{bm}
\usepackage{courier}
\usepackage{graphics}
\usepackage{float}
\usepackage{amsmath}
\usepackage{amssymb}
\usepackage[normalem]{ulem} 
\usepackage[mathscr]{euscript}
\usepackage{acronym}
\usepackage{float}
\usepackage[caption=false]{subfig}
\usepackage{lipsum}
\usepackage{mathtools}
\usepackage{multirow}
\usepackage{graphicx}
\usepackage{mathtools}
\usepackage{multirow}
\usepackage{graphicx}
\usepackage[usenames,dvipsnames,table,xcdraw]{xcolor}
\usepackage[colorlinks, pdfborder={0 0 0}]{hyperref} 
\usepackage[nameinlink,capitalise]{cleveref}
\usepackage{tensor}
\usepackage{verbatim}
\usepackage{wrapfig}
\usepackage{soul} 

\graphicspath{{fig}}

\date{\today}

\begin{document}
\title{Merger-ringdown consistency: A new test of strong gravity using deep learning}

\author{Swetha Bhagwat}
\email{swetha.bhagwat@roma1.infn.it}
\affiliation{%
Dipartimento di Fisica, “Sapienza” Università di Roma \& Sezione INFN Roma1, Roma 00185, Italy
}
\author{Costantino Pacilio}
\email{costantino.pacilio@uniroma1.it}
\affiliation{%
Dipartimento di Fisica, “Sapienza” Università di Roma \& Sezione INFN Roma1, Roma 00185, Italy
}
\begin{abstract}
The gravitational waves emitted during the coalescence of binary black holes are an excellent probe to test the behaviour of strong gravity. In this paper, we propose a new test called the \emph{merger-ringdown consistency test} that focuses on probing the imprints of the dynamics in strong-gravity around the black-holes during the plunge-merger and ringdown phase. Furthermore, we present a scheme that allows us to efficiently combine information across multiple ringdown observations to perform a statistical null test of GR using the detected BH population. We present a proof-of-concept study for this test using simulated binary black hole ringdowns embedded in the next-generation ground-based detector noise. We demonstrate the feasibility of our test using a deep learning framework, setting a precedence for performing precision tests of gravity with neural networks.
\end{abstract}
\maketitle
\section{Introduction}
\label{sec:intro}
The detection of gravitational waves (GWs) emitted during the binary black hole (BH) mergers presents us with an unparalleled opportunity to test the behaviour of strong gravity around BHs \cite{Abbott:2020jks,Nitz_2020}. GWs are emitted as the BHs slowly spiral in towards the common center of mass (a.k.a.~the inspiral phase); this is followed by a rapid plunge and merger (a.k.a.~the plunge-merger phase) where the two BHs coalesce forming a remnant BH which then rings down and settles to a final state (a.k.a.~the ringdown phase) \cite{Maggiore:1900zz}. While both plunge-merger and ringdown contain imprints of the dynamics in the strong field regime at few times the horizon-length scale, the plunge-merger is dictated by non-linear dynamics and the ringdown is prescribed by linear perturbation theory \cite{Kokkotas:1999bd,Konoplya:2011qq,Berti:2009kk}.

Ringdown corresponds to the evolution of linear perturbations on the space-time metric of the remnant \cite{Pani:2013pma}. Given an underlying theory of gravity, the dynamics in the strong-field regime that sets up the perturbation conditions for ringdown and the properties of the remnant BH are not independent. If GR were to be modified in the strong non-linear regime, one would expect the relative excitation of modes in ringdown as well as the final BH's mass and spin to be altered \cite{Kamaretsos:2012bs,Hughes:2019zmt}. We propose a novel test that checks if the excitation conditions set during the plunge-merger phase are consistent with the properties of the remnant BH formed after the ringdown phase. The proposed test checks for the consistency by simultaneously using the frequency content and the amplitudes and phases of excitation in the ringdown signals. Furthermore, we stack the information from multiple GW observations efficiently to provide a statistical `null’ test across a population of binary BH ringdowns. Henceforth, we call it \emph{the merger-ringdown consistency test}. 

We present a complementary test to the already existing battery of tests of GR. Although we draw our inspiration from the IMR test, the two tests address conceptually different questions: while IMR test checks for global consistency of the binary BH evolution \cite{Ghosh:2016qgn,TheLIGOScientific:2016src,Ghosh:2017gfp,Abbott:2020jks}, violation of the merger-ringdown test indicates GR-modifications that alter the perturbations setup for ringdown in a way that is inconsistent with the expected radiated angular momentum and energy in a binary BH coalescence.  However, note that such GR modifications (depending on the details of how GR is modified) might also leave imprints on the global evolution of the binary BH signals, and could be picked up by the IMR test. Furthermore, the merger-ringdown test aims at increasing the sensitivity to the strong field dynamics by zooming in solely on the ringdown phase. Comparing the performance of the two tests in distinguishing GR from non-GR signals is non-trivial and depends on the class of modifications in consideration.

Our test is particularly suited for the third-generation (3g) detectors such as the Einstein Telescope (ET) \cite{Maggiore:2019uih}, the Cosmic Explorer \cite{reitze2019cosmic} and LISA \cite{Audley:2017drz}, where the ringdowns are expected to be loud and the number of detections can be $\sim 10^{3} - 10^{4}/$year \cite{Berti:2016lat,Bhagwat:2016ntk}. Performing prognostic and realistic benchmarking studies on a large number of events with full Bayesian parameter estimation demands for a rapid and computationally efficient inference algorithm. To this aim, we demonstrate the feasibility of our test entirely using a deep learning framework to speed up the parameter inference by orders of magnitude \cite{Green:2020hst,Gabbard:2019rde,Yamamoto:2020rse}. We train a neural network architecture called a conditional variational autoencoder (CVAE) \cite{kingma2013auto,doersch2016tutorial,Erdogan:vae} to infer posterior distributions of the parameter set $\{M,\chi_{f},q\}$ from a set of simulated ringdown waveforms. Following the deep learning application to GW science --- e.g., detection \cite{Gabbard:2017lja,George:2016hay,George:2017pmj,Iess:2020yqj,Wei:2020ztw} and parameter estimation (PE) \cite{Shen:2019vep,Chua:2018woh,Chua:2019wwt,Gabbard:2019rde,Green:2020hst,Green:2020dnx},\footnote{See \cite{Cuoco:2020ogp} for a recent review.} our work also sets a precedence for precision tests of GR using neural networks. Finally, we also demonstrate that deep learning techniques can be efficiently used for population studies for current and next-generation GW detectors.
\section{Merger-ringdown consistency test}
\label{sec:test-itself}
\subsection{Theory}
\label{sec:theory}
Ringdown is modelled as a linear superposition of damped sinusoids with characteristic BH frequencies ($f_{lm}$) and damping times ($\tau_{lm}$) known as the quasi-normal-mode (QNM) spectrum. It is generally decomposed in spin-2 weighted spheroidal harmonic basis $\mathcal{Y}^{lm} (\iota)$, where $(\iota\in[0,\pi))$ is the inclination angle. Ringdowns take the following analytical form \footnote{For simplicity, we decompose the ringdown signal on to spin-2 weighted spherical harmonics basis instead of the more natural spin-2 weighted spheroidal harmonics basis. This assumption is reasonable as long as the spins are not too high and can be estimated following \cite{Berti_2014}.}
\begin{subequations}
\label{eq:waveform}
\begin{align}
& h_+(t)=\frac{M}{D_{L}}\sum_{l,m>0}\mathcal{Y}^{lm}_{+} (\iota)A_{lm}e^{-t/\tau_{lm}} \cos (2\pi f_{lm}t-\phi_{lm}),
\\
 & h_\times(t)=\frac{M}{D_{L}}\sum_{l,m>0}\mathcal{Y}^{lm}_\times(\iota)A_{lm}e^{-t/\tau_{lm}}\sin(2\pi f_{lm}t-\phi_{lm}).
\end{align}
\end{subequations}

Here $\{ +, \times \}$ are the GW polarizations and $D_{L}$ is the luminosity distance of the system. The QNMs are indexed by the angular multipole numbers $(l,m)$ and they are determined by the final mass and final spin $\{M,\chi_{f}\}$, i.e., $f_{lm}=f_{lm}(M,\chi_{f})$ and $\tau_{lm}=\tau_{lm}(M,\chi_{f})$. $A_{lm}$ and $\phi_{lm}$ are the amplitudes and the phases of excitations of QNMs. For a non-spinning binary, the initial system is completely characterized by the total-mass $M_{tot}$ and the binary mass ratio $q$. While $M_{tot}$ sets the overall amplitude scale, $q$ determines the relative excitations of QNMs, i.e., $A_{lm}/A_{22}=A^{R}_{lm}(q)$ and $\phi_{22} - \phi_{lm}=\delta \phi_{lm}(q)$. Thus, the ringdown waveform  can be parameterized by a set of three parameters $\{M, \chi_{f}, q \}$ and Eq.~\eqref{eq:waveform} can be re-written as
\begin{equation}
\label{eq:fitsparam}
h_+(t) = h_+(t; M, \chi_{f}, q)\,,\quad h_{\times}(t) = h_{\times}(t; M, \chi_{f}, q)\,.
\end{equation}
Using the ringdown phase of the GW event one can infer $\{M, \chi_{f}, q \}$ by treating them as independent quantities in  a Bayesian PE setup. 

Next, in GR, a given set of $\{ M_{tot},q\}$ can be deterministically mapped to $\{ M,\chi_{f}\}$ for a non-spinning binary BH system.\footnote{The remnant BH can be expressed in terms of the initial binary BH parameters by fitting the numerical relativity simulations. The relationship can be expressed explicitly in approximate analytical forms \cite{Barausse:2009uz,Pan:2011gk,Barausse:2012qz,Hofmann:2016yih,Husa:2015iqa,Jimenez-Forteza:2016oae,Healy:2016lce}, or implicitly using machine learning algorithms \cite{Varma:2018aht,Haegel:2019uop,Varma:2019csw}.} The three ringdown parameters $\{M, \chi_{f}, q \}$ are not truly independent and, in particular, we map $\chi_f$ to $q$ using the fitting formula presented in \cite{Pan:2011gk} (see also \cite{Barausse:2009uz,Barausse:2012qz,Hofmann:2016yih})
\begin{equation}
\label{eq:fits:spin}
\chi_f=2\sqrt{3}\eta-3.871\eta^2+4.028\eta^3+\mathcal{O}(\eta^3)
\end{equation}
where $\eta=q/(1+q)^2$.

The test checks if the independent measurements of $\{M, \chi_{f}, q \}$ from the ringdowns are consistent with the relation between $\chi_f$ and $q$ as predicted by GR. Specifically, we check if the $\chi_{f}$ directly measured from the ringdown agrees with the $\chi_{f}$ calculated by plugging the measured value of $q$ in Eq.~\eqref{eq:fits:spin}.
\subsection{Prescription for the merger-ringdown consistency }
\label{sec:pres}
Let a population of non-spinning binary BHs ringdowns be detected by a GW observatory. Note that the quantities directly \emph{measured} in PE have a superscript \emph{\rm `meas’} and those \emph{inferred} using Eq.~\eqref{eq:fits:spin} have \emph{\rm `infer’}.

\begin{enumerate}
\item Parametrize the ringdown as in Eq.~\eqref{eq:fitsparam} and estimate $\{M^{\rm meas}, \chi_{f}^{\rm meas}, q^{\rm meas}\}$ for each event. We used the median of the marginalized posterior distribution as the `measured' value.  
\item For each event, compute the $\chi_{f}^{\rm infer}$ from the median value of $q^{\rm meas}$ in step 1 using the relation in Eq.~\eqref{eq:fits:spin}.
\item Make a scatter plot with $\{\chi^{\rm infer},\chi^{\rm meas}\}$ using all ringdown observations. In GR, one expects that all the data should lie along the $\chi^{\rm infer}=\chi^{\rm meas}$ line in a 2-D scatter plot, with the noise in the data leading to a spread around this line. To perform the merger-ringdown consistency test, we express
\begin{equation}
\label{eq:ols}
\chi_f^{\rm meas}=a+b\,\chi_F^{\rm infer}
\end{equation}
and fit for the parameters $\{a, b\}$. If the best-fit parameters for Eq.~\eqref{eq:ols} are compatible with $\{a=0, b=1\}$ the observations are consistent with GR, providing a statistical null test. 
\end{enumerate}

\subsection{Details of Implementation}
\label{sec:our-implementation}
For simplicity, we restrict our study to non-spinning quasi-circular binary BHs. We compute the QNM spectra $ \{f_{lm}(M,\chi_{f}), \tau_{lm}(M,\chi_{f}) \}$ using the data in  \cite{Berti:data}. Further, we focus our attention on the dominant mode $(l,m)=(2,2)$ and the two most excited subdominant angular modes for the case of non-spinning systems --- $(l,m)=\{(2,1),(3,3)\}$ \cite{Bhagwat:2017tkm,Forteza:2020hbw}. We concentrate solely on the dominant overtone, i.e., $n_{\mathrm{overtone}}=0$ \cite{Kamaretsos:2011um,Bhagwat_2020}. We use these simplifying assumptions for this proof-of-concept study. However, note that including more angular modes and overtones is a tangible extension to our work. 

Key to our study is the expression of the QNM excitation amplitudes and phases, as functions of $q$. Following the prescription in \cite{Forteza:2020hbw}, we express $A^{R}_{lm}$ and $\delta \phi_{lm}$ as
\begin{subequations}
\label{eq:fits}
\begin{align}
&A^{R}_{lm}(q)=a_0+\frac{a_1}{q}+\frac{a_2}{q^2}+\frac{a_3}{q^3}\,,\\
&\delta \phi(q)=b_0+\frac{b_1}{b_2+q^2}\,,
\end{align}
\end{subequations}
where we use the convention in which $q>1$. An updated list of coefficients $ \{a_i, b_i \}$ is provided in the Supplemental Material. Further, for the dominant mode’s amplitude, we use $A_{22}=0.86\eta$ \cite{Gossan:2011ha}. Lastly, we assume a uniform support in $\phi_{22} \in [0,2\pi]$ and generate the waveforms expressed in Eq.~\eqref{eq:waveform}.
\section{Deep Learning Framework}
\label{sec:dl}
We use a deep learning framework to reconstruct the posteriors for the parameters $\{M,\chi_f,q\}$ from the waveform. We follow \cite{Gabbard:2019rde,Green:2020hst,Yamamoto:2020rse} and train a CVAE, a neural network architecture well suited to posterior sampling.
\subsection{Details on the CVAE implementation}
\label{sec:methods}
\label{eq:methods}
The CVAE acts as an inverse nonlinear map from the ringdown strain $h=y(x)$ to the posteriors of $x=\{M,\chi_f,q\}$.
\begin{equation}
    \mathrm{CVAE}: y\to p(x|y)\,.
\end{equation}
It has the structure of a variational autoencoder \cite{kingma2013auto,doersch2016tutorial}: it is made of two serial neural network units (the `encoder' and the `decoder') separated by a stochastic latent layer. The first neural network (encoder) maps the input $y$ into the latent layer. The second neural network (decoder) maps the latent representation of the input into the output probability distribution $p(x|y)$.

The CVAE is trained by introducing a third neural network unit, called the auxiliary encoder or the `guide', at training time \cite{tonolini2020variational}. The training consists in optimising a loss function. For the CVAE, the loss naturally splits into \cite{tonolini2020variational, Gabbard:2019rde}:
\begin{enumerate}
    \item the Kullback-Leibler (KL) divergence $\mathcal{L}_{\rm KL}$, measuring the similarity between the outputs of the encoder and the guide; it quantifies the ability of the encoder to produce a meaningful mapping of the input into the latent space;
    \item the reconstruction loss $\mathcal{L}_{\rm recon}$, measuring the probability that the true values $x_{\rm true}$ falls within the decoder distribution.
\end{enumerate}
The total loss to be optimised is
\begin{equation}
    \label{eq:loss}
    \mathcal{L}_{\rm tot}=\mathcal{L}_{\rm recon}+\beta\mathcal{L}_{\rm KL}\,.
\end{equation}
When $\beta=1$, $\mathcal{L}_{\rm tot}$ coincides with the standard ELBO loss \cite{kingma2013auto,doersch2016tutorial}. The additional parameter $\beta$ gives flexibility in implementing effective training strategies.

\begin{figure*}
    \centering
    \includegraphics{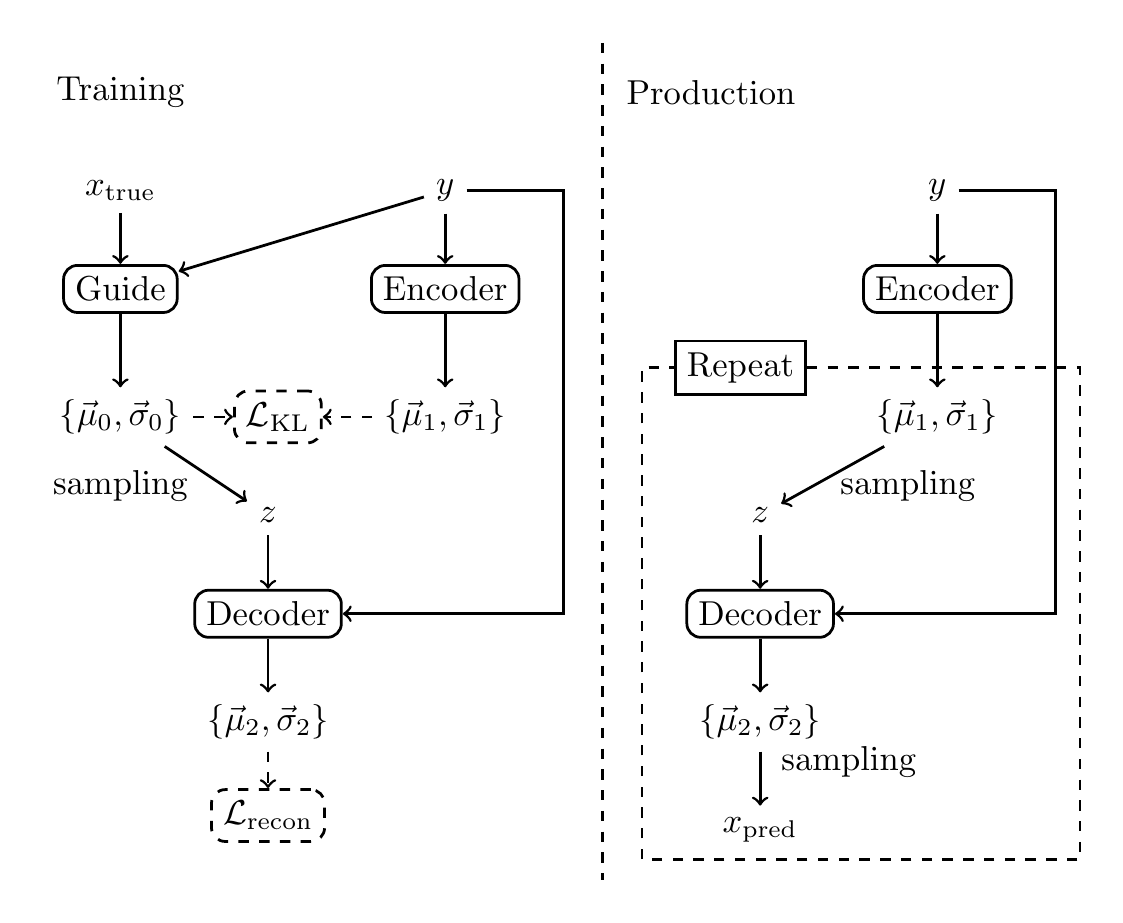}
    \caption{A schematic representation of the CVAE architecture. On the left, a single training step is represented. First, the signal $y$ is mapped by the encoder into a latent stochastic distribution, which is a multivariate diagonal Gaussian with means and standard deviations $\{\vec{\mu}_1,\vec{\sigma}_1\}$; similarly, the couple $(y,x_{\rm true})$ is mapped by the guide into a second Gaussian with parameters $\{\vec{\mu}_0,\vec{\sigma}_0\}$; the two are then combined into the $\mathcal{L}_{\rm KL}$ loss. Next, a latent variable $z$ is sampled from the guide distribution and mapped by the decoder into a third Gaussian with parameters $\{\vec{\mu}_2,\vec{\sigma}_2\}$. This final distribution is eventually used to compute the $\mathcal{L}_{\rm recon}$ loss. On the right, a single step at production time is shown. Now, the latent representation is sampled from the encoder and a predicted output $x_{\rm pred}$ is sampled from the decoder; this step is repeated $n_{\rm samples}$ times to produce an informative posterior distribution for $x$; in text, we fix $n_{\rm samples}=10^4$. Note the final distribution of $x_{\rm pred}$ is not a Gaussian, but is a complex distribution resulting from the convolution of the two serial sampling steps.}
    \label{fig:cvae}
\end{figure*}

After the training, the guide is dropped out. At production time, only the encoder and the decoder are used to sample the posteriors $p(x|y)$. Fig.~\ref{fig:cvae} contains a flow-diagram of the training and production steps. More details on the neural network architecture and our codes are provided in a dedicated \texttt{git} repository \cite{mrt:2020}.

The hyperparameters which determine the CVAE training are listed in Tab.~\ref{tab:hyperparams}.
\begin{table}[]
    \centering
    \begin{tabular}{l|c}
        \hline\hline
        \\[-1em]
         Batch size & $512$\\
         Epochs & $500$\\
         Optimizer & Adam\\
         Initial lr & $10^{-4}$\\
         lr decay & $\times0.5$ every $80$ epochs\\
         $\beta$ annealing & $3\times[10^{-5},\frac{1}{3},\frac{2}{3},1,1,1]$\\
         Validation fraction & $10\%$\\
         \hline\hline
    \end{tabular}
    \caption{Hyperparameters used for training the CVAE.}
    \label{tab:hyperparams}
\end{table}
We train the CVAE in batches of $512$ waveforms for $500$ epochs, i.e., $500$ forward-backward passes of the entire training set. The loss is minimized using the Adam optimizer with an initial learning rate set to $10^{-4}$. The learning rate is decreased by a factor of $2$ every $80$ epochs. To monitor the convergence of the loss, we set aside $10\%$ of the training dataset and we use it for validation. Following \cite{Green:2020hst}, we use $\beta$ to implement a cyclic annealing schedule. Annealing improves the efficiency of the training and allows autoencoders to express more meaningful latent variables \cite{fu2019cyclical}. We increase $\beta$ from $0$ to $1$ in steps of $[10^{-5},\frac{1}{3},\frac{2}{3},1,1,1]$ and these steps are repeated $3$ times. After this, $\beta$ is definitively fixed to 1.

Next, the training performances improve when the inputs $y$ are standardized to zero mean and unit variance, and when the outputs $x$ are normalized to have support in $[1,100]$. $x$ is then scaled back to the original normalization at production time.
\subsection{Network training}
\label{sec:training}
To train the network, we simulate a dataset of $10^5$ ringdowns by sampling the waveform parameters uniformly in the ranges indicated in Tab.~\ref{tab:parameters:choice}. The ringdown waveforms are sampled at $4096$ Hz with a total signal duration of $31.25$ ms, thus corresponding to arrays of length $128$. Signal-to-noise ratio (SNR) is used to set the waveform scale w.r.t.~the noise. The SNR is computed as in \cite{Berti:2005ys,Baibhav:2018rfk}. When performing PE, we only estimate posteriors for $\{M,\chi_f,q\}$.
\begin{table}[h]
    \centering
    \begin{tabular}{l|c|c}
        Parameter & Symbol & Range \\
        \hline\hline
        \\[-1em]
         Final BH mass & $M$& $[25,100]~M_\odot$\\
         Final BH spin & $\chi_f$ & $[0,0.9]$\\
         Binary mass ratio & $q$ & $[1,8]$\\
         Phase of the (2,2) mode & $\phi_{22}$ & $[0,2\pi]~{\rm rad}$\\
         Signal-to-noise ratio & SNR & $[40,80]$\\
         \hline\hline
    \end{tabular}
    \caption{Ranges for the waveform parameters. All the parameters are sampled uniformly. Note we marginalize over the last two parameters.}
    \label{tab:parameters:choice}
\end{table}
For simplicity, we only consider the $+$ polarization and fix the inclination angle to $\iota=\pi/3$. The ringdowns are embedded in simulated ET-like noise segments \cite{Hild:2010id,ET:sensitivities}. At each training iteration, we assign noise instances randomly to the waveforms to prevent the CVAE from learning spurious correlations between the waveforms and the noise realizations.

\begin{figure}
    \centering
    \includegraphics[width=0.42\textwidth]{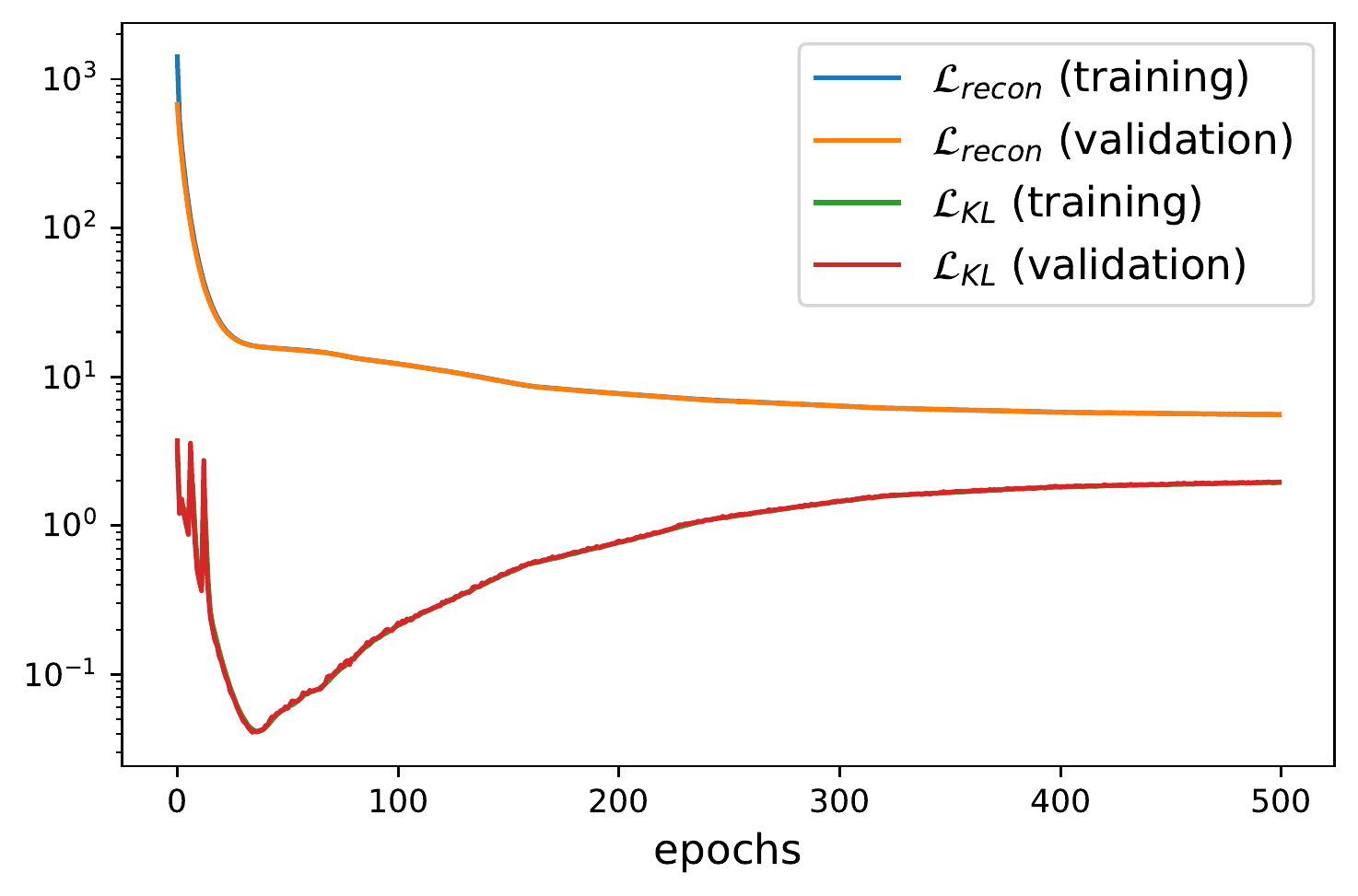}
    \caption{Evolution of the reconstruction and KL losses across the training epochs.}
    \label{fig:losses}
\end{figure}

Our training takes $84$ minutes on a single GPU. Fig.~\ref{fig:losses} shows the evolution of the reconstruction loss $\mathcal{L}_{\rm recon}$ and the KL divergence $\mathcal{L}_{\rm KL}$ separately. Further, we show the loss evaluated on the $90\%$ training dataset and on the $10\%$ validation dataset. Notice that the training and validation losses are consistent, substantiating that the network is not overfitting.\footnote{The initial oscillations which are visible in $\mathcal{L}_{\rm KL}$ are due to the cyclic annealing.}%

\begin{figure}
    \centering
    \includegraphics[width=0.46\textwidth]{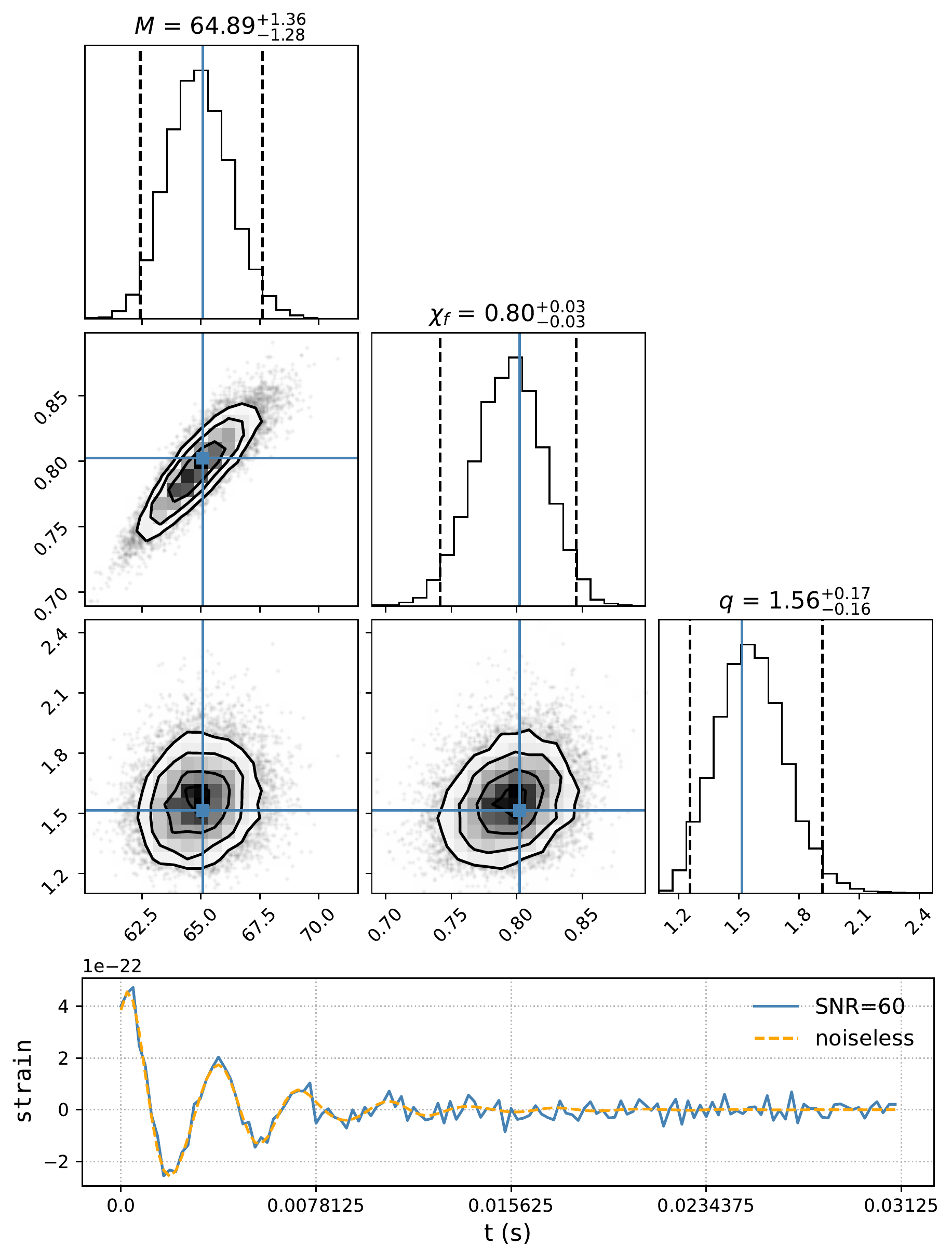}
    \caption{Contour plot for the PE of the ringdown signal shown in the lower panel, with $(M,\chi_f,q)=(65,0.8,1.5)$ and SNR$=60$. Blue lines indicate the true values. Dashed lines mark the $95\%$ confidence interval. The plot titles indicate $1\sigma$ uncertainties.}
    \label{fig:contour}
\end{figure}

\begin{figure}
    \centering
    \includegraphics[width=0.4\textwidth]{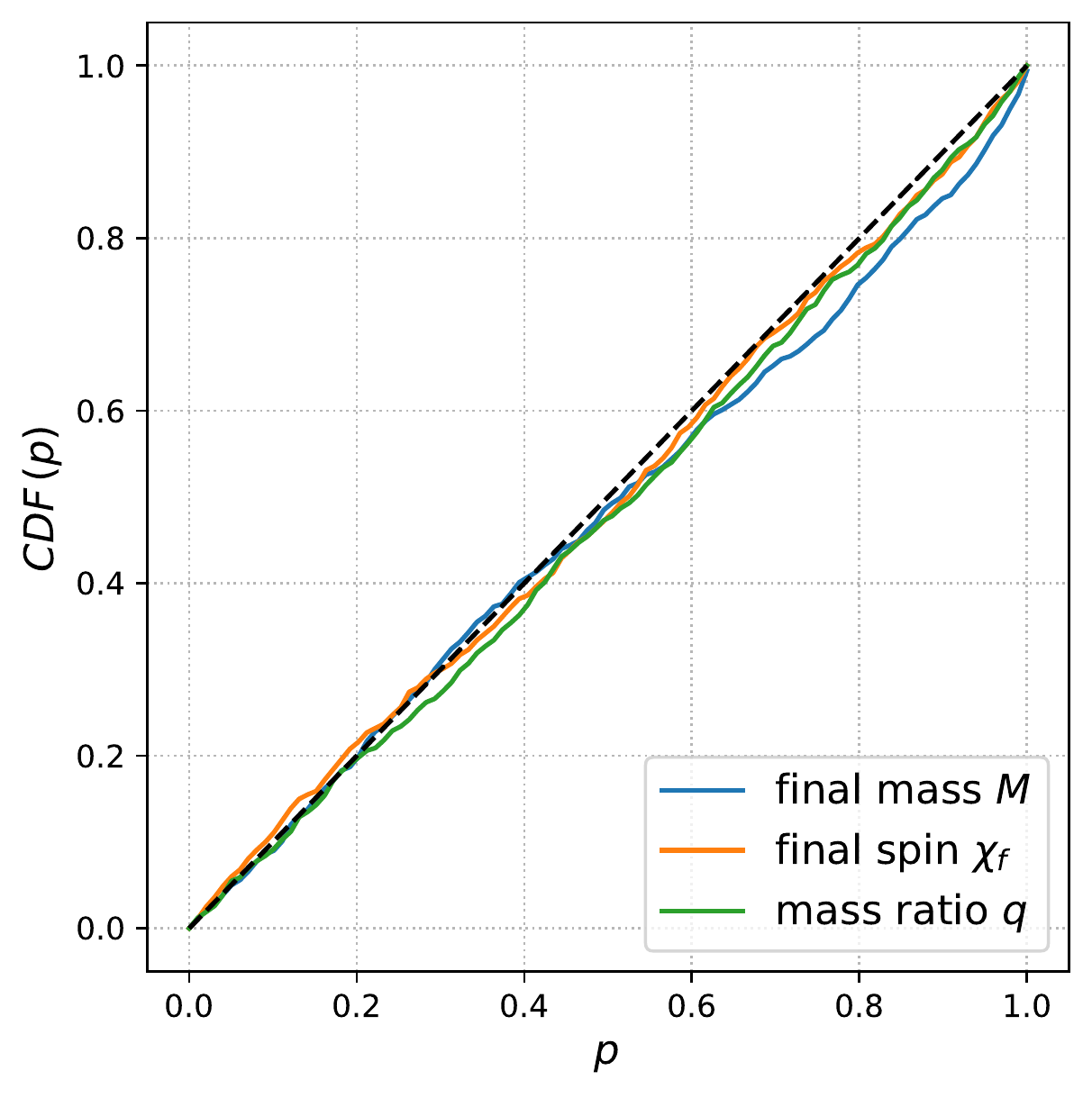}
    \caption{P-P plot of test dataset. For each observable, the plot shows the cumulative distribution CDF of the true values as a function of the $p\,\%$ confidence interval of the posterior distribution.}
    \label{fig:pp}
\end{figure}

To test the network, we generate a new dataset of $10^3$ simulated ringdown waveforms, whose parameters are sampled again from the ranges indicated in Tab.~\ref{tab:parameters:choice}. For each input waveform, the CVAE samples $10^4$ distinct points to build the posterior. The total time to analyze all the samples is approximately $40$ s on a single GPU, corresponding to $40$ ms per waveform. For illustration, Fig.~\ref{fig:contour} shows the contour plot obtained from the PE of one signal from the test dataset.
For a quantitative diagnostic of the network performances, we present the P-P plot in Fig.~\ref{fig:pp}: the plot shows marginalized cumulative distribution CDF of the true values $x_{\rm true}$ as a function of $p\,\%$ confidence interval. A diagonal P-P plot means that $x_{\rm true}$ is contained $p\,\%$ of the times within $p\,\%$ confidence interval of the marginalized posteriors for $x_{\rm pred}$. Note that all the CDFs are consistent with the diagonal, demonstrating that our CVAE recovers the posterior samples for $\{M,\chi_f,q\}$ from the ringdowns reliably.
\section{Results}
\label{sec:results}
We present our proof-of-concept study in two parts. In section \ref{sec:test} we demonstrate that when our test is applied to a set of ringdowns consistent with GR, these signals satisfy the null test. Next, in section \ref{sec:test-nGR} we use 4 sets of non-GR ringdowns and show that with $\sim 20-50$ events the non-GR signals violate our null test - allowing us to distinguish GR from non-GR ringdown.
\subsection{Proof-of-concept Merger-Ringdown test: GR signals}
\label{sec:test}
We simulate a dataset consisting of $10^3$ ringdowns with parameter ranges as presented in Tab.~\ref{tab:parameters:choice} except for $\chi_f$. Here, $\chi_f$ is inferred from $q$ by imposing the relation \eqref{eq:fits:spin}. 

\begin{figure}
    \centering
    \includegraphics[width=0.4\textwidth]{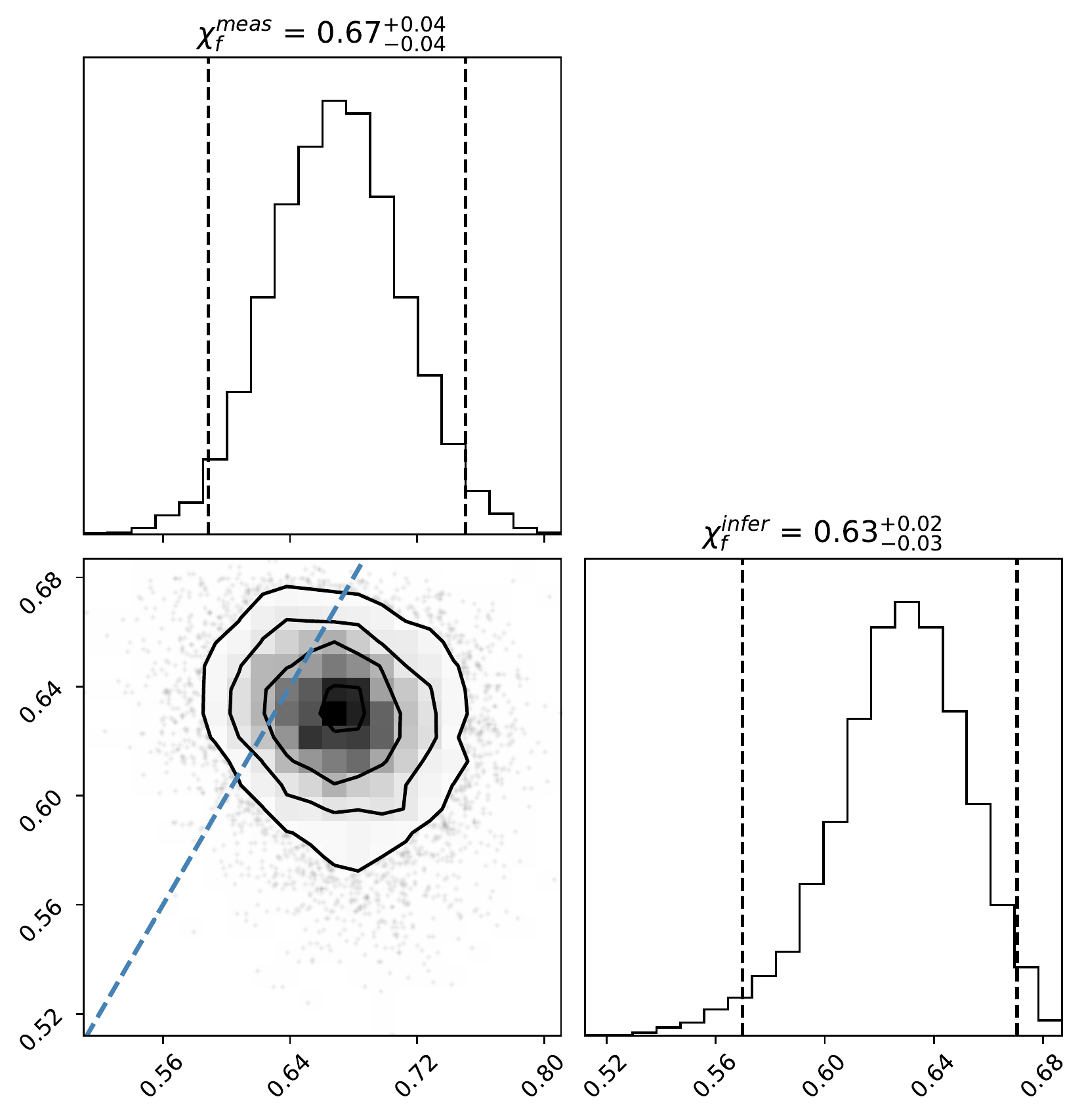}
    \caption{Contour plot for the posteriors of $\chi_f^{\rm meas}$ and ~$\chi_f^{\rm infer}$ for a signal with $(M,\chi_f,q)=(52.6,0.66,1.53)$ and SNR$=60$. The signal is extracted from the second test dataset, where the relation \eqref{eq:fits:spin} is enforced. The blue dashed line represents $\chi_f^{\rm meas}=\chi_f^{\rm infer}$ line.}
    \label{fig:identity:contour}
\end{figure}

First, the posteriors produced by our CVAE for a single event can be used to check if the event validates the spin relation \eqref{eq:fits:spin}. In Fig.~\ref{fig:identity:contour}, we show the posterior samples for $\chi_f^{\rm meas}$ and $\chi_f^{\rm infer}$ --- where $\chi_f^{\rm infer}$ is determined by $q^{\rm meas}$ as per Eq.~\eqref{eq:fits:spin}, for a randomly chosen event from our dataset. For this event, we see that $\chi_f^{\rm meas}=\chi_f^{\rm infer}$ (blue dashed line) lies within the $68\%$ credible interval of the posteriors, asserting that this event is consistent with GR evolution.

\begin{figure}
\centering
\includegraphics[width=0.43\textwidth]{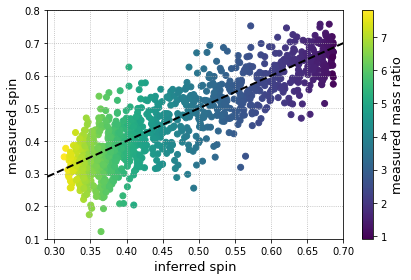}
\caption{Scatter plot of $\chi_f^{\rm meas}$ vs $\chi_f^{\rm infer}$. The color bar indicates the value $q^{\rm meas}$ for each observation. The black dotted line marks the $\chi_f^{\rm meas}=\chi_f^{\rm infer}$.}
\label{fig:scatter}
\end{figure}

Further, we use the scheme outlined in Sec.~\ref{sec:pres} to combine the information across multiple ringdown observations for a more stringent test of GR, as illustrated in Fig \ref{fig:scatter}. For a noiseless GR ringdown, we expect $\chi_f^{\rm meas}=\chi_f^{\rm infer}$. However, our inferences are probabilistic and contain noise. This leads to measurement uncertainties that translate as a scatter around the diagonal line. 

In Fig.~\ref{fig:scatter}, we confirm that our dataset lies around $\chi_f^{\rm meas}=\chi_f^{\rm infer}$. Also, as expected, lower values of $q$ give higher values of $\chi_{f}$. A weighted least squared (WLS) fit for  Eq.~\eqref{eq:ols} gives $a\in[-0.014,0.014]$ and $b\in[0.963,1.013]$ at $95\%$ confidence level, showing an agreement with the $\chi_f^{\rm meas}=\chi_f^{\rm infer}$ line. We weighed each event by $(\sigma_{\chi_f^{\rm meas}}\sigma_{\chi_f^{\rm infer}})^{-1/2}$ to emphasize the more confident recoveries.\footnote{We verified that the results of an ordinary (unweighted) least squared fit do not significantly change.} Fig.~\ref{fig:scatter} thus observationally validates Eq.~\eqref{eq:ols}.

\begin{figure}
\centering
\includegraphics[width=0.43\textwidth]{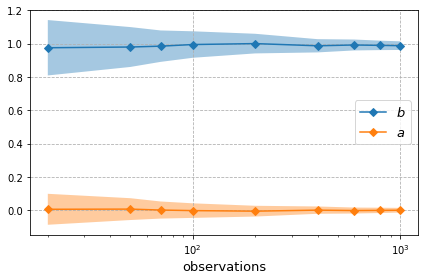}
\caption{Evolution of the best-fit values (continuous lines) and $2\sigma$ contours (shaded regions) for $b$ and $a$ in Eq.~\eqref{eq:ols}, versus the number of observations. The mean values and the confidence intervals 
are averaged over 10 random realisations. In GR, the noiseless best-fit corresponds to $b=1$ and $a=0$.}
\label{fig:ols}
\end{figure}

Next, to assess the efficiency of this test with the number of observations, we study the convergence of $a$ and $b$. In Fig.~\ref{fig:ols}, we present the (weighted) best-fit values for $a$ and $b$ with their $2\sigma$ confidence levels as a function of the number of observations.

In Fig.~\ref{fig:ols}, we find that the mean value for $a$ and $b$ are $\sim 0$ and $1$, respectively, even for a small number of observations. We see that the uncertainties in the measurements of $a$ and $b$, i.e., $\{ \sigma_{b}, \sigma_{a}\}$ decrease with increasing number of observations as a power-law. For instance, with 20 observations we can constrain $\{ \sigma_{b}(n=20), \sigma_{a}(n=20)\} = \{0.0832, 0.00465 \}$ and with 100 observations we have $\{ \sigma_{b}(n=100), \sigma_{a}(n=100)\} = \{0.0398, 0.0219 \}$. Concretely, $\sigma_{a}$ and $\sigma_{b}$ scale with the number of observations as  
\begin{equation}
\label{eq:fits:ols}
\sigma_{a}(n) = \frac{0.21}{\sqrt{n}}\,,\qquad\sigma_{b}(n) = \frac{0.41}{\sqrt{n}}\,,
\end{equation}
which is consistent with our expectations given our noise model.

Thus, while the merger-ringdown consistency test is powerful when combining a large number of ringdowns, it is also feasible to perform it with just a few observations ($\sim 20$). 

\subsection{Proof-of-concept Merger-Ringdown test: non-GR signals}
\label{sec:test-nGR}
In this section, we present the performance of the test on a set of non-GR ringdown signals. We consider phenomenological deviations from GR without assuming physical mechanisms responsible for the GR modifications. Specifically, we generate 4 sets of non-GR ringdowns by heuristically modifying amplitudes and phases of mode excitations
\begin{subequations}
\begin{align}
    &A^R_{lm}\to(1+\Delta A)A^R_{lm,GR}\\
    &\delta\phi_{lm}\to(1+\Delta\delta\phi)\delta\phi_{lm,GR}
\end{align}
\end{subequations}
with the 4 distinct cases enlisted as 
\begin{subequations}
\label{eq:cases}
\begin{align}
    &\text{Case 1:} &\Delta A=0\quad&\Delta\delta\phi=0.1\\
    &\text{Case 2:} &\Delta A=0\quad&\Delta\delta\phi=-0.1\\
    &\text{Case 3:} &\Delta A=0.1\quad&\Delta\delta\phi=0.1\\
    &\text{Case 4:} &\Delta A=0.1\quad&\Delta\delta\phi=-0.1
\end{align}
\end{subequations}

This class of GR-modifications implies an altered relation between the ringdown perturbation conditions and the final properties of the remnant; therefore, we expect our test to be naturally sensitive to these modifications. Furthermore, note that other than the amplitude ratios and phase differences between the modes, the details of the non-GR signal are identical to the GR ones --- i.e., the QNM spectrum is unaltered.

\begin{figure*}[t!]
\includegraphics[width=0.4\textwidth]{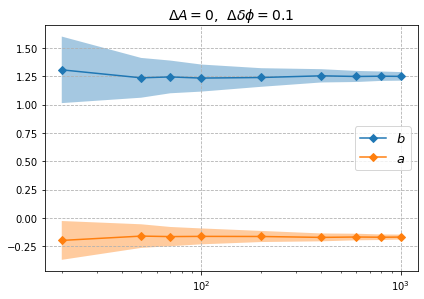}
\includegraphics[width=0.4\textwidth]{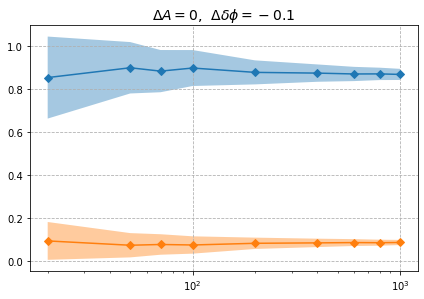} \\
\includegraphics[width=0.4\textwidth]{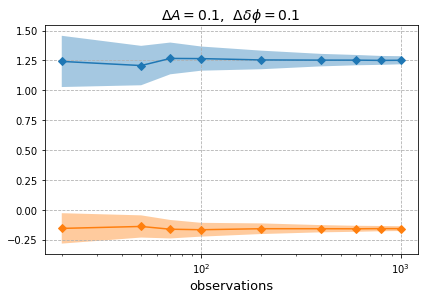}
\includegraphics[width=0.4\textwidth]{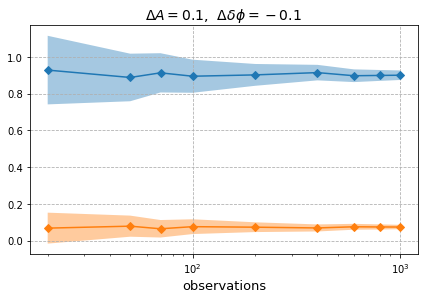}
\caption{Similar to Fig.~\ref{fig:ols} but applied on the set of non-GR test signals described through Case 1-4 in Eq.~\ref{eq:cases} (depicted as panels, clockwise from the top-left). Each set contains $10^3$ signals. The mean values and the confidence intervals are averaged over 10 random realisations. Note that for all of the cases we get best fit $a\ne0$ and $b\ne1$ with $95\%$ credible intervals with less than 50 observations.}
\label{fig:non-GR}
\end{figure*}

The best-fit values of $a$ and $b$ in Eq.~\ref{eq:ols} for each of the non-GR signal-set are presented in Fig.~\ref{fig:non-GR}. We remind that any departure from $a=0$ and $b=1$ indicates that the dataset contains ringdown that do not satisfy the GR null test. We note that the merger-ringdown test successfully excludes the  $a=0$ and $b=1$ for a relatively small number of observations. Specifically, GR values are excluded at a $2\sigma$ level with $\mathcal{O}(20)$ observations for Case 1-3 and with $\mathcal{O}(50)$ observations for Case 4. 
\section{Conclusion and Outlook}
\label{sec:discussion}
We demonstrated a proof-of-concept study for a novel test of GR called \emph{the merger-ringdown consistency test} that checks for statistical consistency between the plunge-merger-ringdown phase across a set of ringdown detections using a deep learning framework. The test aims at increasing the sensitivity to the merger-ringdown by simultaneously incorporating information on both the amplitude-phase excitations and the QNM frequency spectra in the ringdown. It uses the fact that the QNM complex excitation amplitudes and the spectrum in the ringdown are not independent quantities in GR. The test provides an efficient way of stacking ringdowns. Furthermore, for possible future detections of heavy mass binary systems (that is set precedence by the discovery of \cite{Abbott_2019}), our test can be used to check the consistency of initial BH parameters with the ringdown alone if the inspiral is not well measured.

This work illustrates that Bayesian deep learning methods can be applied to infer the posteriors of ringdown parameters to conduct precision tests of GR. Unlike the traditional Monte-Carlo sampling, neural networks perform the PE in fractions of a second, providing a crucial edge when dealing with the large number of observations that are expected with future GW detectors.

We also highlight that, in its most generic construct, the TIGER infrastructure allows to identify parametric deviations from GR in the waveform by applying a Bayesian hypothesis testing \cite{TIGER-concept-1,TIGER-concept-2}. In testing GR with ringdown signals, the most popular strategy has been to parametrize the deviations in the QNM frequencies and damping times, followed by evaluating the Bayes factor for GR versus non-GR hypothesis \cite{Meidam_2014,Gossan:2011ha,Brito:2018rfr}. Our test is conceptually different, because we are not explicitly checking whether the QNM spectrum of the remnant is consistent with that of a general relativistic BH. Rather, we focus on the consistency between the relative amplitudes and phases of the ringdown modes with the spin of the remnant BH. The modifications to which the two tests are sensitive do not necessarily overlap; we expect our test to be efficient in scenarios where the departure from GR influences the relative amplitudes and phase, but the final remnant is similar to a general relativistic BH. Comparing the efficiency of our test to the TIGER implementation is not straightforward and needs further exploration.

Here we have used non-spinning progenitor BHs where the QNM excitation amplitudes and phases are fully parametrizable by its mass ratio $q$ and the $\chi_{f}-q$ relation is approximated by the simple analytical expression in Eq.~\eqref{eq:fits:spin}. However, our method can be extended to encompass spinning progenitor BHs where the QNM excitations depend on both $q$ and $\chi_{1,2}$. The dependence of the remnant spin $\chi_{f}$ on the binary BH parameters should then be replaced by implicit non-analytical relations such as those in \cite{Varma:2018aht,Haegel:2019uop}. Our WLS fit strategy does not rely on analytical relations and new parameters can be estimated by increasing the output dimensions of the CVAE.

This study used stellar-mass BH ringdowns targeting the ET-like data. Similar results can be expected to hold for CE. However, LISA will detect ringdowns from supermassive BHs \cite{Berti:2016lat,Baibhav:2018rfk}, with loud SNRs. We plan to extend our analysis to include LISA-like data in the future. Finally, in this work, we demonstrate the feasibility of a \emph{null} test of GR, by implementing our test on GR as well on a class of phenomenologically constructed non-GR ringdowns. An interesting extension to our work would be an extensive study on non-GR signals in a parameterized framework such as that in ParSpec \cite{Maselli:2019mjd}. 
\begin{acknowledgments}
We thank Paolo Pani and Andrea Maselli for their productive comments on an early version of this manuscript and Guido Sanguinetti for his advices on neural networks. We thank Nathan Johnson-McDaniel for useful clarifications about the IMR test. CP is indebted to Enrico Barausse and Luca Heltai for their invaluable support in getting familiar with neural networks and gravitational wave physics. We acknowledge financial support provided under the European Union's H2020 ERC, Starting Grant agreement no.~DarkGRA--757480. We also acknowledge support under the MIUR PRIN and FARE programmes (GW-NEXT, CUP:~B84I20000100001), and from the Amaldi Research Center funded by the MIUR program ``Dipartimento di Eccellenza'' (CUP: B81I18001170001).
\paragraph*{Software.} The \texttt{PyCBC} \cite{Canton:2014ena,Usman:2015kfa} library was used to generate the noise realizations and \texttt{LalSuite} \cite{lalsuite} to generate the ringdowns. The neural network model used in this work was developed using the \texttt{PyTorch} library \cite{NEURIPS2019_9015} for \texttt{Python}. The WLS test was performed with the \texttt{statsmodels} package \cite{statsmodels} for \texttt{Python}. The corner plots have been produced with the \texttt{corner} package \cite{corner} for \texttt{Python}.
\paragraph*{Code availability.} The code used for this paper is made available in a dedicated \texttt{git} repository \cite{mrt:2020}. 
\end{acknowledgments}
\appendix*
\section{Excitation amplitudes and phases}
We update the fits presented in \cite{Forteza:2020hbw} by additionally requiring $A_{lm}/A_{22} \to 0$ for $q \to 1$ \cite{Kamaretsos:2011um} and present the coefficients for Eq.~\eqref{eq:fits} in Tab.~\ref{tab:fits}. The start of ringdown is chosen at $t_{peak} + 12 M$. Note that the goodness of the fits do not change significantly between the version here and in \cite{Forteza:2020hbw}.
\begin{table}[h!]
\centering
\setlength{\tabcolsep}{1em}
\begin{tabular}{ccc}
& $(3,3)$ & $(2,1)$\\
\\[-1em]
\hline\hline
\\[-1em]
$a_0$ & 0.433253 & 0.472881\\
$a_1$ & -0.555401 & -1.1035\\
$a_2$ & 0.0845934 & 1.03775\\
$a_3$ & 0.0375546 & -0.407131\\
\hline
\\[-1em]
$b_0$ & 2.63521 & 1.80298\\
$b_1$ & 8.09316 & -9.70704\\
$b_2$ & 8.32479 & 9.77376
\end{tabular}
\caption{Values of the fit coefficients in Eq.~\eqref{eq:fits} for $(l,m)=(3,3)$ and $(2,1)$.}
\label{tab:fits}
\end{table}
\newpage
\bibliography{main}
\end{document}